\documentclass[%
 reprint,
 amsmath,amssymb,
 aps,
]{revtex4-2}

\usepackage{graphicx}
\usepackage{dcolumn}
\usepackage{bm}
\usepackage{xcolor}

\begin{document}

\preprint{APS/123-QED}

\title{Inverse cascade in zonal flows.}

\author{Siddhant Mishra}
\author{Anikesh Pal}%
 \email{pala@iitk.ac.in}
\affiliation{%
 Department of Mechanical Engineering\\
Indian Institute of Technology, Kanpur 208016, India
}%




\date{\today}

\begin{abstract}
Zonal winds on Jovian planets play an important role in governing the cloud dynamics, transport of momentum, scalars, and weather patterns. Therefore, it is crucial to understand the evolution of the zonal flows and their sustainability. Based on studies in two-dimensional (2D) $\beta$ plane setups, zonal flow is believed to be forced at the intermediate scale via baroclinic instabilities, and the inverse cascade leads to the transfer of energy to large scales. However, whether such a process exists in three-dimensional (3D) deep convection systems remains an open and challenging question. To explore a possible answer, we perform Large Eddy Simulations at the geophysically interesting regime of $Ra=$$10^{12}$, $Ek=$$10^{-6}$,$10^{-7}$ and $10^{-8}$ in horizontally rotating Rayleigh-B\'enard convection setup and discover the existence of natural forcing through buoyancy and inverse cascade. The turbulent kinetic energy budget analysis and the spectral space assessment of the results corroborate the emanation of a strong mean flow from chaos.
\end{abstract}

\maketitle


Turbulence, jets, waves, and vortices characterize planetary circulation. An essential member of this group is the zonal flow (jets), especially on Joivan atmospheres such as Jupiter and Saturn. Moving along the east-west direction, the strength of zonal flow varies with latitude, but the strongest ones are located near the equatorial region. The strength of the zonal flow depends on the underlying turbulence that generates it. In many turbulent flow cases, the energy is transferred from the large to small scales, however in a geophysical context the large-scale circulations emerge as a consequence of an upscale transfer/inverse cascade in certain 2D quasi-geostrophic limits.
Early studies have shown that in the presence of the geophysical $\beta$ effect, closely packed eddies transform to zonally elongated structures \cite{rhines1975waves}. This hypothesis was extended to account for flows on the Jovian planets under barotropic and baroclinic setups \cite{williams1978planetary,williams1979planetary}. The barotropic representation used implicit forcing to consider the baroclinic effects \cite{williams1978planetary}. Under conditions of extreme barotropization, the flow transitions to zonostrophic turbulence characterized by alternating jets similar to Jupiter and Saturn \cite{galperin2006anisotropic,sukoriansky2007arrest,sukoriansky2008nonlinear}. In contrast, the baroclinic setup revealed that the jets always exist with ambient eddies \cite{williams1979planetary,panetta1993zonal}, a phenomenon common on Jovian atmospheres, especially on Jupiter. Further, studies have also explored the effect of topography and large-scale circulations under the baroclinic setup \cite{thompson2010jet,boland2012formation,chen2013effects,berloff2011latency}.  Barotropic and baroclinic models emphasized that zonal winds are manifestations of shallow layers forced by baroclinic instabilities and relied significantly on two-dimensional quasigeostrophic approximation. Simultaneously the deep convection model also became popular \cite{busse1976simple,busse1994convection}. According to this model, a barotropic interior, caused by the homogenization of entropy gradients due to interior convection, allows the formation of Taylor columns. The non-linear interaction of these columns then generates equatorial and high-latitude zonal winds \cite{galperin2019zonal}. Mathematically, this involves solving the three-dimensional governing differential equations with either the anelastic approximation \cite{jones2009linear,gastine2012effects,heimpel2016simulation} or the Boussinesq assumption\cite{aurnou2001strong,christensen2001zonal,christensen2002zonal,heimpel2005simulation,nicoski2024asymptotic} and rotation. The key governing parameters in such a system include Rayleigh number($Ra$), Ekman number($Ek$), and Prandtl number($Pr$). The setup can be further simplified in terms of geometry to a cartesian domain, a setup famously known as the Rayleigh-B\'enard convection (RBC) \cite{ahlers2009heat}.\\

Zonal flow is the mean horizontal component when large-scale shear develops in RBC \cite{goluskin2014convectively}. The associated shear of zonal flow disrupts the convective structures to reduce heat transport. Simulations of zonal flow in 2D RBC are the limiting case of $Ek \rightarrow 0$ and neglect the effect of finite rotation. Recently, 3D RBC setups with rotation along the horizontal axis have demonstrated the development of strong zonal winds \cite{von2015generation}. Similar to the 2D studies \cite{goluskin2013zonal,goluskin2014convectively}, zonal flow in 3D forms after the convective rolls annihilate. With horizontal rotation the flow resembles those near the equator in a typical $f$-plane setup \cite{pedlosky2013geophysical} and longitudinal variation of the Coriolis force is excluded.\\

The use of RBC with horizontal rotation to study the generation of zonal flow or convective rolls has gained attraction recently \cite{von2015generation,ludemann2022transition,Ludemann2023-nu,wang2020zonal}. These studies have focused on different aspects of horizontal rotation such as the aspect ratio \cite{wang2020zonal} and elliptical instability in the convective roll phase \cite{ludemann2022transition}. However, the literature on 3D zonal flows in horizontally rotating RBC is limited to moderate $Ra$ of $10^7$ and $10^{8}$. To date, it remains an open question on the formation and evolution of zonal flows at higher Ra and rotation rates relevant to geophysically interesting regimes. We attempt to answer this question by performing large eddy simulations (LES) of horizontally rotating RBC over the parameters $Ra=$$10^{12}$ and $Ek=$$10^{-6}$,$10^{-7}$  and $10^{-8}$ and $Pr=$$0.71$. LES provides a suitable alternative to DNS \cite{von2015generation, wang2020zonal, wang2021influence} to reach very high $Ra$ while using a relatively coarse grid. Some existing work has utilized LES for simulating RBC without rotation \cite{kimmel2000large, peng2006large} but the use of LES for rotating RBC remains unreported. We study the transition of the convective rolls to zonal flow without imposing any shear profile as an initial condition \cite{wang2020zonal}. The evolution of the Reynolds number based on the mean kinetic energy and the turbulent kinetic energy demonstrates the advent of a zonal flow, further implying a gain of energy in the mean scales at the expense of turbulence. We calculate the total kinetic energy spectra and the energy flux on different horizontal planes and conclude that buoyancy forces the small scales on the horizontal planes from where it is transferred upscale. Many researchers have also investigated the energy and its flux in the spectral space for the RBC to study the upscale transfer and the inverse cascade mechanism. However, these studies created barotropic and baroclinic modes by taking averages along the vertical. In our study, we perform spectral analysis on the full three-dimensional velocity field to study the upscale transfer of energy and provide evidence for the inverse cascade mechanism \cite{alexakis2024large,de2024pattern}.\\

We consider the buoyant convection of an incompressible Boussinesq fluid between a hot (bottom) and cold (top) plate (see Fig S1 in SI). The temperature difference between the plates is constantly maintained at $\Delta T$, and the system is subjected to background rotation \textit{\textbf{$\Omega$}} = $\Omega$$\hat{e_{2}}$ about the horizontal axis, perpendicular to gravity given by \textit{\textbf{g}} = $-g$$\hat{e_{3}}$. The fluid is Newtonian with density $\rho$, kinematic viscosity $\nu$, thermal diffusivity $\kappa$, and adiabatic volume expansion coefficient $\alpha$. We select the vertical height $H$, the temperature difference $\Delta T$, and free-fall velocity $u_{f}=\sqrt{g\alpha \Delta T H}$ as the scales for length, temperature, and velocity to non-dimensionalize the governing equations \cite{pandey_2018,naskar_2022a,naskar_2022b}. The size of the domain is $L_{1} = L_{2} = 2\pi$ in the horizontal direction and $L_{3} = H = 1$ in the vertical direction. The filtered non-dimensional governing equation \cite{pham2014large} for large eddy simulations takes the following form
\begin{equation}\label{Eq:cont}
    \frac{\partial \overline{u}_{j}}{\partial x_{j}} = 0,
\end{equation}
\begin{equation}\label{Eq:NSeq}
\begin{split}
    \frac{\partial \overline{u}_{i}}{\partial t} +
    \frac{\partial \overline{u}_{j}\overline{u}_{i}}{\partial x_{j}}= -\frac{\partial \overline{P}}{\partial x_{i}}+ \frac{2}{Ek}\sqrt{\frac{Pr}{Ra}}\epsilon_{ij2}\overline{u}_{j}+\\
    \overline{\theta}\delta_{i3}+
    \sqrt{\frac{Pr}{Ra}}\frac{\partial^2 \overline{u}_{i}}{\partial x_{j} \partial x_{j}}
    -\frac{\partial \tau_{ij}}{\partial x_{j}},
\end{split}
\end{equation}
\begin{equation}\label{Eq:energ}
    \frac{\partial \overline{\theta}}{\partial t} + \frac{\partial \overline{u}_{j}\overline{\theta}}{\partial x_{j}} = \frac{1}{\sqrt{RaPr}}\frac{\partial^2 \overline{\theta}}{\partial x_{j} \partial x_{j}} -
    \frac{\partial Q_{j}}{\partial x_{j}}.
\end{equation}
Here the $\overline{...}$ denotes the filtered quantities. The additional terms $\tau_{ij}$ and $Q_{j}$ represent the sub-grid stress and sub-grid buoyancy flux arising out of the filtering operation (see SI for details). In the vertical direction, we use impenetrable free-slip plates with Dirichlet boundary condition on $\overline{\theta}$ ($\pm=0.5$ at $x_{3}=$$\mp 0.5$) whereas the horizontal directions are considered periodic (see SI). We use the dynamic Smagorinsky model proposed by \cite{germano1991dynamic,lilly1992proposed}. The dynamic model has been successfully used to simulate different problems \cite{tejada2007langmuir,armenio2002investigation,taylor2007internal,pham2014large} relevant to geophysical flows. We include more details in the SI section on LES, boundary conditions, and validation (Fig S2). \\

The governing equations have three non-dimensional parameters: (i) Rayleigh number($Ra$), the measure of buoyancy force to viscous force (ii) Ekman number($Ek$), the ratio of viscous to Coriolis force and (iii) Prandtl number($Pr$), the measure of momentum to thermal diffusivity. The mathematical forms of these parameters are given in SI. We perform simulations for different sets of parameters (see Table S1 in SI) such that the cases can be categorized according to $1/Ro = 1.6852$, $16.852$, and $168.52$ ($Ro$ is the Rossby number, see SI for details). Geophysically horizontal rotation models the flow in $f$-plane setup \cite{pedlosky2013geophysical}, where the latitude is fixed to 0$^{\circ}$, resembling flow at the equator. Further, \cite{novi2019rapidly} used this $f$-plane setup to model flows at different latitudes. Under this $f$-plane setup, winds moving in the $+x_{1}$ direction are referred to as prograde (eastward) and $-x_{1}$ as retrograde (westward) to relate it to directions of wind movement on different planets.\\

\begin{figure*}
\begin{center}
\includegraphics[width=1.0\linewidth]{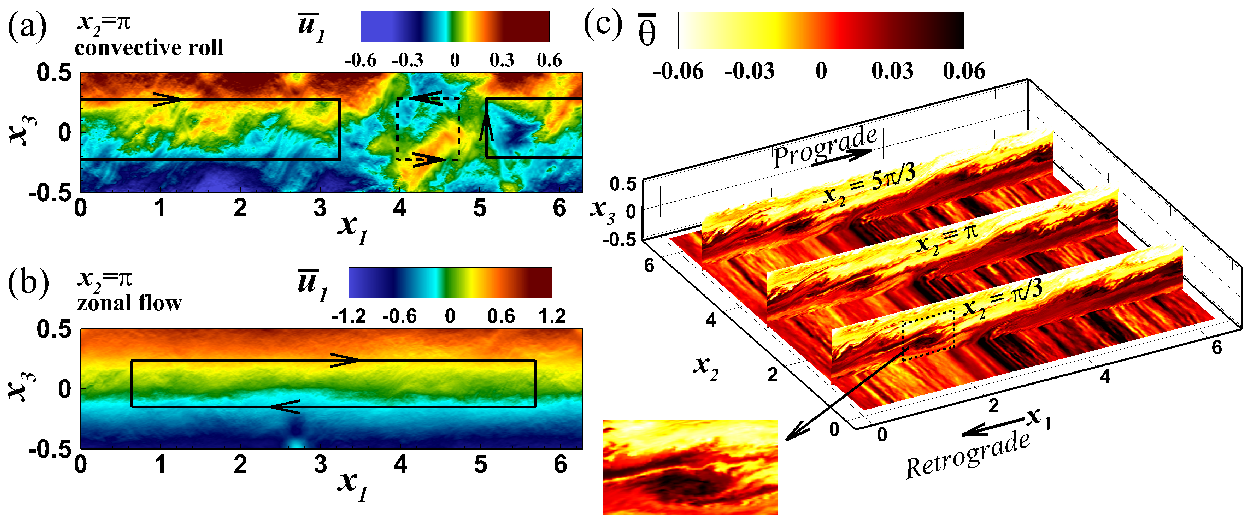}
\end{center}
\caption{Instantaneous visualization of the convective roll and zonal flow states for $Ra=$$10^{12}$ and $Ek=$$10^{-7}$. In (a) and (b) the solid arrows denote cyclonic roll and dashed arrows represent anti-cyclonic roll.} 
\label{Fig1}
\end{figure*}

We demonstrate the convective roll phase for $Ra = 10^{12}$ and $Ek = 10^{-7}$ using the horizontal velocity field $\overline{u_{1}}$ in figure \ref{Fig1}(a). These convective rolls are characterized by a widening cyclonic roll and a narrowing anticyclonic roll \cite{von2015generation}.  
The zonal flow state is initiated when the anti-cyclonic roll completely disappears. Figure \ref{Fig1}(b) shows the zonal flow has $\overline{u_{1}} > 0$ near the top plate, $\overline{u_{1}} < 0$ near the bottom and is vertically sheared. The impact of this shear is visible on the thermal plumes as shown using the $\overline{\theta}$ contours at the vertical planes $x_{2}$ = $\pi$/3, $\pi$ and 5$\pi$/3 in figure \ref{Fig1}(c). The hot thermal plume gets sheared by the retrograde ($- x_{1}$) wind near the bottom plate and the prograde ($+ x_{1}$) winds shear the cold plumes near the top plate, thus establishing large-scale vertically sheared flow with a significant mean in the horizontal direction (see supplementary movies S1 and S2). We observe similar flow evolution for other Rayleigh and Ekman numbers (see Fig S3 in SI, supplementary movies S3, S4, S5, and S6). Such large-scale shear with horizontal mean components has been studied in a 2D RBC setup \cite{goluskin2013zonal,goluskin2014convectively,fitzgerald2014mechanisms,liu2024fixed}. The strength and cause of this mean flow in 3D horizontally rotating RBC in the parameter space ($Ra-Ek$), however, remain unexplored \cite{von2015generation,wang2020zonal}.\\

\begin{figure*}
    \begin{center}
        \includegraphics[width=1.0\linewidth]{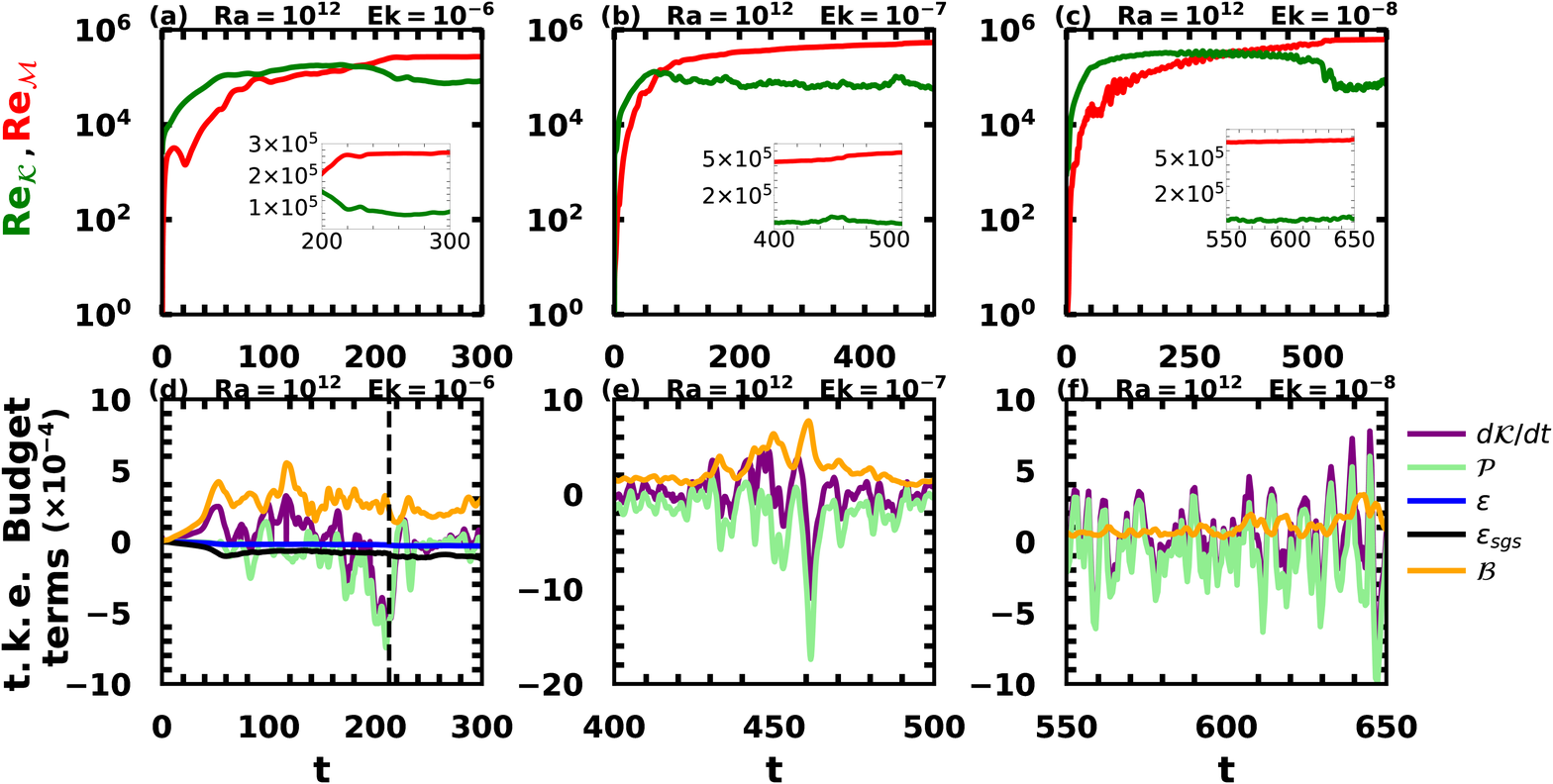}
    \end{center}
    \caption{$Re_{\mathcal{K}}, Re_{\mathcal{M}}$((a)-(c)) and $t.k.e.$ budget((d)-(f)) plots with time.The $t.k.e.$ budget plots in (e) and (f) are shown for the time as per the inset of (b) and (c) respectively. The time, t in the plots is non-dimensional.}
    \label{Fig2}
\end{figure*}

To quantify the strength of the mean shear, we define the mean kinetic energy in terms of the Reynolds number $Re_{\mathcal{M}}$. As we will demonstrate that this mean emerges at the expense of the turbulent kinetic energy $(t.k.e.)$, we also define $Re_{\mathcal{K}}$ (see SI). Figures \ref{Fig2}(a) - \ref{Fig2}(c) shows the variation of $Re_{\mathcal{K}}$ and $Re_{\mathcal{M}}$ for different parameters. In figure \ref{Fig2}(a) for $Ra=$$10^{12}$ and $Ek=$$10^{-6}$, we see that during the convective roll phase $Re_{\mathcal{K}} < Re_{\mathcal{M}}$. However, when $Re_{\mathcal{M}}$ overtakes $Re_{\mathcal{K}}$ the zonal flow phase starts, where the mean becomes stronger than the turbulence. We observe a similar pattern for the parameters $Ra=$$10^{12}$,$Ek=$$10^{-7}$ (figure \ref{Fig2}b) and $Ra=$$10^{12}$,$Ek=$$10^{-8}$ (figure \ref{Fig2}c). The insets in figure \ref{Fig2}b and \ref{Fig2}c show the quantities far away from the point from where the mean overtakes turbulence. To further enhance our understanding of how this exchange between the mean and turbulence is facilitated we analyze the turbulent kinetic energy ($t.k.e.$, see equations $14$-$19$ in SI) budget terms as shown in figures \ref{Fig2}(d)-\ref{Fig2}(f). The $t.k.e.$ budget equation signifies energy conservation. We achieve a closure of this equation which ensures proper resolution to resolve the relevant scales of motion. We see that at $Ra=$$10^{12}$ and $Ek=10^{-6}$, during the convective roll phase the buoyancy flux ($\mathcal{B}$) converts the potential energy to $t.k.e.$. At this instant, a portion of this $t.k.e.$ is transferred to the mean field as shown by the negative production term while the remaining $t.k.e.$ dissipates owing to $\epsilon_{sgs}$ and $\epsilon$. $\mathcal{P} > 0$ represents extraction and energy transfer from the mean to turbulence. In contrast $\mathcal{P} < 0$ indicates energy transfer from small-scale turbulence to mean fields \cite{kundu2015fluid}. Note that $\epsilon_{sgs} \ge \epsilon$ implies the utility of our LES model for Rayliegh-B\'enard convection at high $Ra$. When $Re_{\mathcal{M}}$ overtakes $Re_{\mathcal{K}}$ in figure \ref{Fig2}(a), we find a sudden drop in $\mathcal{P}$ to negative values implying energy draining from turbulence to the mean.  The flow transitions in this instance from convective rolls to zonal flow (see supplementary movies S3 and S4). With the increase in the rotation rates to $Ek=$$10^{-7}$ and $10^{-8}$ the energy transfer from turbulence to mean becomes significant as shown in figures \ref{Fig2}(e) and \ref{Fig2}(f) respectively. Note that we only show the $t.k.e.$ budget terms after the formation of zonal flow for $Ek=$$10^{-7}$ and $10^{-8}$ as shown in the insets of figures \ref{Fig2}(b) and \ref{Fig2}(c) respectively. This significant energy transfer for $Ek=$$10^{-7}$ and $10^{-8}$ compared to $Ek=$$10^{-6}$, implies stronger zonal flow. In many turbulent flows, an imposed anisotropic mean field facilitates turbulence production \cite{pham2014evolution,pham2014large,olsthoorn2023dynamics}. However, in the present scenario, the mean anisotropy naturally develops from the underlying buoyant small-scale turbulence owing to horizontal rotation.\\ 

\begin{figure*}
    \begin{center}
    \centering
        \includegraphics[width=1.0\linewidth]{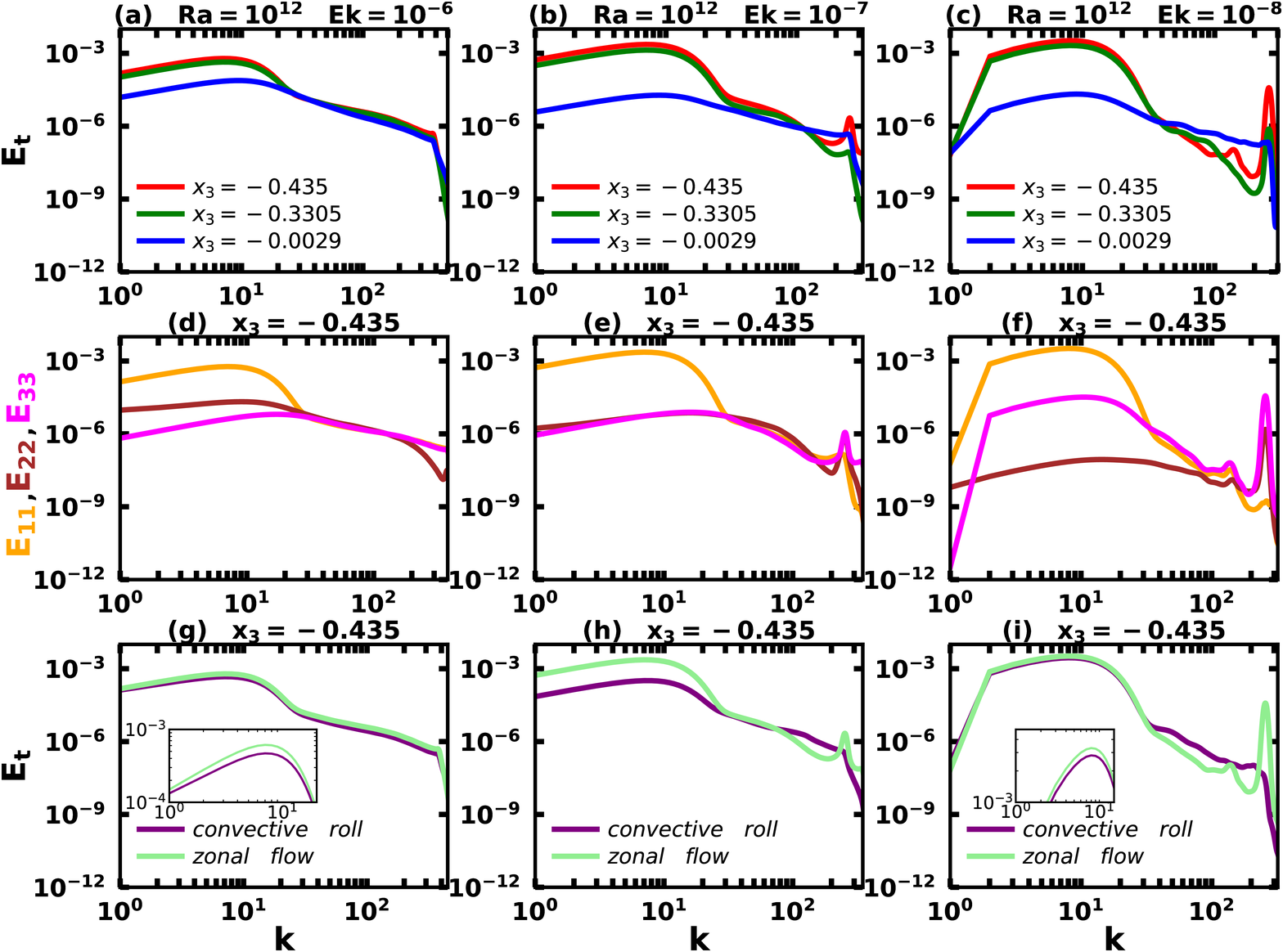}
    \end{center}
    \caption{The total kinetic energy spectra plots for  $Ra=$$10^{12}$,$Ek=$$10^{-6}$(left panel), $Ra=$$10^{12}$,$Ek=$$10^{-7}$(middle panel) and $Ra=$$10^{12}$,$Ek=$$10^{-8}$(right panel). The top and the middle row (a-f) are the spectra at instances when zonal flow has already been established. The bottom row (g-i) compares the spectra at the convective roll phase and the zonal flow state.}
    \label{Fig3}
\end{figure*}

We plot the total kinetic energy spectra (see SI) in figure \ref{Fig3} to further corroborate the energy transfer from small scales to the mean field. The left panel shows the spectra plots for $Ra=$$10^{12}$,$Ek=$$10^{-6}$, middle panel for $Ra=$$10^{12}$,$Ek=$$10^{-7}$ and right panel for $Ra=$$10^{12}$, $Ek=$$10^{-8}$. These plots provide an estimate of energy distribution across different scales of motion. For different parameters (figures \ref{Fig3}(a)-\ref{Fig3}(c)), we find a sudden jump in the energy spectra toward the edge. The magnitude of this sudden jump decreases as we move up vertically from $x_{3}=$$-0.435$ (near the wall) towards $x_{3}=$$-0.0029$ (middle of the domain). Several studies have investigated the effect of forcing and subsequent cascade of energy in two-dimensional turbulence \cite{yakhot1999two,boffetta2012two,poujol2020role,sofiadis2023inducing} and more recently in three-dimensional turbulence \cite{alexakis2024large,de2024pattern}. Unlike the literature, in our cases, the forcing at the small scales is natural due to the buoyant term in the momentum equation \cite{goluskin2013zonal}. The source of the thermal forcing is the thermal plumes that lose coherence while proceeding from $x_{3}=-0.435$ to $x_{3}=-0.0029$, owning to dispersion caused by the sheared velocity field. Figures \ref{Fig3}(d)-\ref{Fig3}(f) show the contribution to $E_{t}(k)$ from the components $E_{11}$, $E_{22}$ and $E_{33}$. The forcing is predominantly in the $E_{33}$ spectra rather than $E_{11}$. Note that $E_{11}$ stays at least two orders of magnitude above $E_{33}$ or $E_{22}$ at the large scales. The buoyancy forces the components along the gravity, from where the anisotropic effect of rotation causes it to transfer to zonally extended component $E_{11}$ via non-linear momentum transfer \cite{goluskin2013zonal,von2015generation}. Further, the plots in figures \ref{Fig3}(g)-\ref{Fig3}(i) show that energy in the zonal flow state is greater than the convective roll phase at large scales. It can also be observed from the right panel in figure \ref{Fig3} that the extent of forcing at the small scales increases with an increase in rotation. Therefore, we conclude that rotation significantly influences the transfer of energy to $E_{11}$.\\

\begin{figure*}
    \begin{center}
        \includegraphics[width=\linewidth]{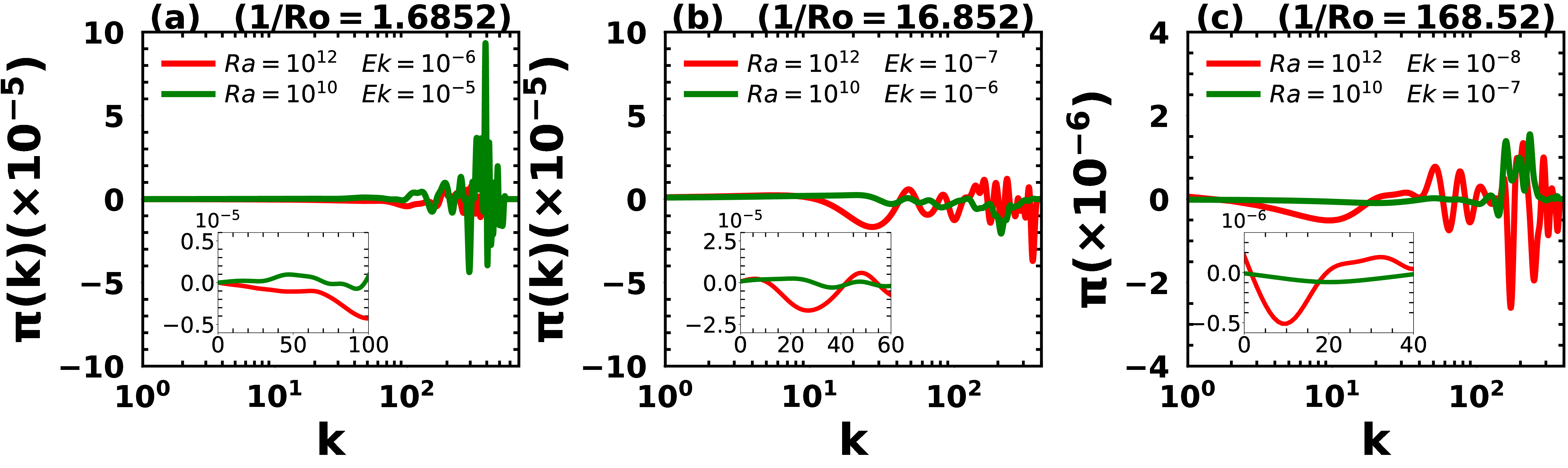}
    \end{center}
    \caption{The energy flux $\Pi(k)$ for 1/Ro = (a) 1.6852 (b) 16.852 and (c) 168.52. The inset shows the range of $k$ for which $\Pi(k)$ remains invariably negative.}
    \label{Fig4}
\end{figure*}

The advection term in the momentum equation is responsible for the energy transfer across different scales of motion. The upscale energy transfer in zonal flows has been suggested earlier \cite{heimpel2005simulation,goluskin2013zonal} but has never been proved for zonal flows. Therefore, we compute the energy flux, $\Pi(k)$ (see SI) in figure \ref{Fig4} to demonstrate the upscale energy transfer. Since the thermal forcing is most significant near the boundaries, we compute $\Pi(k)$ on the plane $x_3=-0.435$. Intuitively, $\Pi(k)$ could be understood as the counterpart of $\mathcal{P}$ in the wavenumber domain. $\Pi(k) > 0$ implies the transfer of energy to small scales (forward cascade) and the energy moves towards large scales when $\Pi(k) < 0$ (upscale/inverse cascade) \cite{alexakis2018cascades}. As shown in figure \ref{Fig4}, the thermal forcing causes scale-to-scale fluctuation of $\Pi(k)$ near the forcing wavenumbers (see figure \ref{Fig3}) for all the cases. The insets in figures \ref{Fig4}(a)-\ref{Fig4}(c) focus on the range of $k$ for which $\Pi(k)$ remains invariably negative. We see from these insets that with an increase in the rotation rate, as indicated by increasing $1/Ro$, the range of $k$ with invariably negative energy flux shifts towards larger length scales. This also indicates that faster rotation energizes the larger scales resulting in stronger zonal flow. The non-linear exchanges carry the energy upscale at these small wavenumbers (large scales) after being thermally forced at large wavenumbers (small scales). We also find that the magnitude of $\Pi(k)$ in the region of upscale transfer is less for $Ra=$$10^{10}$ compared to $Ra=$$10^{12}$, signifying that higher thermal forcing facilitates the upscale energy transfer at large scales.\\

Large eddy simulations are performed in a three-dimensional horizontally rotating Rayleigh-B\'enard convection setup at geophysically interesting regime ($Ra=10^{12}$,$Ek=$$10^{-6}$,$10^{-7}$ and $10^{-8}$) to elucidate the formation of zonal flows. We find that zonal flows have strong mean components that develop at the expense of the underlying turbulence, and analyze the $t.k.e.$ budget to understand the energy exchange between the mean flow field and turbulence. The buoyancy flux ($\mathcal{B}$) feeds energy to turbulence establishing a convective roll phase. The convective rolls grow in size and finally merge to form zonal flows. At this instance, the mean kinetic energy overtakes the $t.k.e.$ and the production ($\mathcal{P}$) becomes negative signifying the energy transfer from the small-scale turbulent field to the mean flow.\\ 

The analysis of total kinetic energy spectra (figure \ref{Fig3}) and energy flux (figure \ref{Fig4}) demonstrates that zonal flow is forced by buoyancy at small scales. The kinetic energy cascades upscale to small wavenumbers via the non-linear interactions \cite{heimpel2005simulation,von2015generation}. Inverse cascade has previously been investigated in vertically rotating Rayleigh-B\'enard convection\cite{rubio2014upscale,favier2014inverse} by decomposing the flow into baroclinic
(depth-dependent) and barotropic modes (depth-independent). The energy supplied by the baroclinic modes transfers to the large scales in the barotropic mode. The process results in a feedback loop where large baroclinic scales take energy out of the barotropic mode and smaller ones deposit energy back to the barotropic mode. Our work reports an upscale transfer without any such decomposition. Our methodology is more general similar to the recent approaches to studying large-scale self-organization \cite{alexakis2024large} and pattern formation\cite{de2024pattern}. External forcing in barotropic setups to mimic the baroclinic effects is a common practice, thus, the spectra reported in our work resemble those reported in the barotropic setups \cite{sukoriansky2007arrest,sukoriansky2008nonlinear}. However, we strongly emphasize that we develop strong zonal flows without two-dimensional approximations or the $\beta$ effect.\\

We report an upscale transfer of energy in horizontally rotating Rayleigh-B\'enard convection and provide evidence of inverse cascade in zonal flows by analyzing the horizontal $x_{1}-$$x_{2}$ planes. The presence of non-periodic vertical boundaries restricts us from performing the Fourier transform in the vertical direction (see SI). Nonetheless, we capture the natural forcing at the small scales resulting in the upscale energy transfer. Unlike the investigations of inverse cascade in a triply periodic domain that requires external forcing at scales equivalent to domain vertical height \cite{alexakis2018cascades}, our system is forced naturally through buoyancy at small scales and provides a feasible explanation of the mystery of the zonal flow at the Jovian planets. A possible extension of the present work will be to investigate the upscale energy transfer in horizontally rotating Rayleigh-B\'enard convection with anelastic approximations \cite{heimpel2016simulation}. The imposition of a mean or a varying magnetic field in horizontally rotating Rayleigh-B\'enard convection can also lead to a change in flow structures, and it will be intriguing to quantify the energy transfer and partition between the mean and turbulent components of the velocity and magnetic fields.\\


%

\end{document}